\documentclass[11pt]{article}
\usepackage{amsmath}
\usepackage{graphicx}
\usepackage[usenames,dvipsnames,table]{xcolor}
\usepackage{amssymb}
\usepackage{float}
\usepackage{scrtime}
\usepackage{fancyhdr}
\usepackage{subfigure}
\usepackage{authblk}
\usepackage{lipsum}
\usepackage{rotating}
\usepackage{floatpag}
\usepackage{varioref}
\usepackage{slashed}
\usepackage{enumerate}
\usepackage{listings}
\usepackage{cancel}
\usepackage{wrapfig}

\usepackage{array}
\makeatletter
\newcommand{\thickhline}{%
    \noalign {\ifnum 0=`}\fi \hrule height 1pt
    \futurelet \reserved@a \@xhline
}
\newcolumntype{"}{@{\hskip\tabcolsep\vrule width 1pt\hskip\tabcolsep}}
\makeatother

\usepackage{caption}
\DeclareCaptionFont{white}{\color{white}}
\DeclareCaptionFormat{listing}{%
  \parbox{\textwidth}{\colorbox{gray}{\parbox{\textwidth}{#1#2#3}}\vskip-4pt}}
\usepackage[colorlinks=true
,urlcolor=blue
,anchorcolor=blue
,citecolor=blue
,filecolor=blue
,linkcolor=black
,menucolor=blue
,pagecolor=blue
,linktocpage=true
]{hyperref}

\usepackage[numbers,sort&compress]{natbib}

\setlipsumdefault{131}

\newcommand{\code}[1]{{\small{\texttt{#1}}}}

\newcommand{\ra}{\ensuremath{\rightarrow}}

\newcommand{\SO}[1]{\ensuremath{\mathrm{SO}(#1)}}
\newcommand{\SU}[1]{\ensuremath{\mathrm{SU}(#1)}}

\newcommand{\bs}[1]{\ensuremath{\boldsymbol{#1}}}

\def\etmiss{E^\text{miss}_\text{T}}
\def\pt{p_\text{T}}

\def\tanb{\text{tan}\;\beta}

\newcommand{\half}{\frac{1}{2}}

\labelformat{section}{Section #1}
\labelformat{subsection}{Section #1}
\labelformat{equation}{Eq.~(#1)}
\labelformat{figure}{Fig.~#1}
\labelformat{subfigure}{Fig.~\thefigure#1}
\labelformat{table}{Tab.~#1}
\labelformat{lstlisting}{Listing~#1}

\newcommand{\mc}[3]{\multicolumn{#1}{#2}{#3}}

\def\s{\sigma}

\def\a{\alpha}

\def\be{\begin{equation}}
\def\ee{\end{equation}}
\def\bea{\begin{eqnarray}}
\def\eea{\end{eqnarray}}

\graphicspath{{./figures/}}

\begin{document}

\floatpagestyle{plain}

\pagenumbering{roman}

\renewcommand{\headrulewidth}{0pt}
\rhead{
OHSTPY-HEP-T-14-002\\
}
\fancyfoot{}

\title{\huge \bf{LHC Phenomenology of\\ \bs{\SO{10}} Models with Yukawa
Unification II}}

\author[$\dag$]{Archana Anandakrishnan}
\author[$\dag$]{B. Charles Bryant}
\author[$\dag$]{Stuart Raby}

\affil[$\dag$]{\em Department of Physics, The Ohio State University,\newline
191 W.~Woodruff Ave, Columbus, OH 43210, USA \enspace\enspace\enspace\enspace
\medskip}

\maketitle
\thispagestyle{fancy}

\begin{abstract}\normalsize\parindent 0pt\parskip 0pt
In this paper we study Yukawa-unified SO(10) SUSY GUTs with two types of SO(10) boundary conditions: (i) universal gaugino masses and (ii) non-universal gaugino masses with \emph{effective ``mirage''} mediation. With these boundary conditions, we perform a global $\chi^2$ analysis to obtain the parameters consistent with 11 low energy observables, including the top, bottom, and tau masses. Both boundary conditions have universal scalar masses and ``just so'' splitting for the up- and down-type Higgs masses. In these models, the third family scalars are lighter than the first two families and the gauginos are lighter than all the scalars. We therefore focus on the gluino phenomenology in these models. In particular, we estimate the lowest allowed gluino mass in our models coming from the most recent LHC data and compare these to limits obtained using simplified models. We find that the lower bound on $M_{\widetilde{g}}$ in Yukawa-unified SO(10) SUSY GUTs is generically $\sim$1.2 TEV at the $1\s$ level unless there is considerable degeneracy between the gluino and the LSP, in which case the bounds are much weaker. Hence many of our benchmark points are not ruled out by the present LHC data and are still viable models which can be tested at LHC 14. 
\end{abstract}

\clearpage
\newpage

\pagenumbering{arabic}

\section{Introduction}

Supersymmetric grand unified theories (SUSY GUTs) have the remarkable feature of gauge coupling unification~\cite{Dimopoulos:1981yj,Dimopoulos:1981zb, Ibanez:1981yh,Sakai:1981gr, Einhorn:1981sx, Marciano:1981un}. This supplies an experimental clue for weak scale SUSY but does not significantly constrain the sparticle spectrum. It has been observed however that SUSY GUT models which have third family Yukawa coupling unification, such as SO(10) or 
$\SU{4}_c \times \SU{2}_L \times\SU{2}_R$, can place considerable constraints on the low energy SUSY 
spectrum~\cite{Blazek:2001sb, Blazek:2002ta, Baer:2001yy, Tobe:2003bc}. The choice of boundary conditions for the soft terms at the GUT scale determines these constraints and can lead to dramatically different low energy SUSY spectra. Requiring Yukawa unification and obtaining good fits to top, bottom, and tau masses limits the number of viable boundary 
conditions~\cite{Blazek:2002ta, Baer:2009ie, Badziak:2011wm, Anandakrishnan:2012tj}. It is therefore manageable to survey the phenomenological implications for the LHC of viable Yukawa-unified SUSY GUTs. We restrict our studies to two types of SO(10) boundary conditions: (i) universal gaugino masses and (ii) non-universal gaugino masses with \emph{effective ``mirage''} mediation~\cite{Anandakrishnan:2013cwa}. Both boundary conditions have universal scalar masses and ``just so'' splitting for the up- and down-type Higgs masses. In these models, the third family scalars are lighter than the first two families~\cite{Pierce:1996zz}. Additionally, the gauginos are lighter than all the scalars. We therefore focus on the gluino phenomenology in these models.

Exclusion limits for the gluino are reported in the context of various Simplified Models by the ATLAS and CMS collaborations. The current exclusion limits on simplified models in which the gluino decays to a pair of third family quarks and the lightest neutralino or chargino are $\sim$1.3 TeV at the $1\s$ 
level~\cite{ATLAS-CONF-2013-061,CMS-PAS-SUS-13-004}. In Ref.~\cite{Anandakrishnan:2013nca}, the phenomenological consequences of the Yukawa-unified SO(10) SUSY GUT with universal gaugino masses were explored using three different sparticle searches performed by the CMS 
collaboration~\cite{Chatrchyan:2013wxa, Chatrchyan:2012paa, Chatrchyan:2013lya}. A lower bound of $\sim$1 TeV was found for the gluino mass, which is $\sim$20\% below the bound placed by simplified models in the searches considered.\footnote{Unfortunately, the branching ratios of the gluino presented in Ref.~\cite{Anandakrishnan:2013nca} had an error due to the manner in which our spectrum calculator \code{maton} was interfaced with \code{SDECAY} (which calculated the decay branching ratios). The error affected the specific branching ratios reported in Tab. 3 of~\cite{Anandakrishnan:2013nca}, but the overall results quoted in the paper remain unchanged.} The results reported in this paper supersede those reported in Ref.~\cite{Anandakrishnan:2013nca}. In addition, we also study the exclusion limits for the gluino in models with non-universal gaugino masses. To complement the CMS searches, we evaluate each of the benchmark models with the code \code{CheckMATE}~\cite{Drees:2013wra}, which is optimized for simulating many ATLAS searches. We present here the results for the exclusion limits on the gluino mass in the two variants of Yukawa-unified SO(10) SUSY GUTs.

In Section 2 we present the benchmark models derived for the different boundary conditions considered. The relevant experimental analyses and results are presented in Section 3. We give concluding remarks in Section 4.

\section{Benchmark Models}
A global $\chi^2$ analysis was performed by varying the GUT scale model parameters, listed in~\ref{tab:parameters}, to fit 11 low energy observables, $M_W$, $M_Z$, $G_F$, $\alpha_{em}^{-1}$, $\alpha_s(M_Z)$, $M_t$, $m_b(m_b)$, $M_{\tau}$, $\mathcal{B}(B\rightarrow X_s \gamma)$, $\mathcal{B}(B_s \rightarrow \mu^+ \mu^-)$, and $M_h$~\cite{Anandakrishnan:2012tj,Anandakrishnan:2013cwa}. Good fits to the low energy observables determined the SUSY spectrum in two scenarios: universal and non-universal gaugino masses. In Ref~\cite{Anandakrishnan:2012tj}, a full three family model was also considered with an additional $D_3$ family symmetry and good fits to 36 observables were obtained, but it was shown that the soft SUSY spectrum was completely determined by considering a third family model and fitting the observables listed above. For the purposes of discussing the LHC phenomenology of these models, we will limit the analysis in this paper to the third family model. Note that by good fits we refer to $\chi^2/d.o.f.$ less than 1.0, 2.6, and 3.8 at 68\%, 90\%, and 95\% confidence levels, respectively. The number of degrees of freedom ($d.o.f.$) is defined to be the difference between the number of observables used in the fit and the number of parameters that are allowed to vary in the analysis.

\begin{table}[h!]
\begin{center}
\renewcommand{\arraystretch}{1.2}
\scalebox{0.83}{
\begin{tabular}{|l||c|c||c|c||}
\hline
Sector &  Universal Gaugino Masses & \# & Non-universal Gaugino Masses& \# \\
\hline
gauge             & $\alpha_G$, $M_G$, $\epsilon_3$                   & 3  &
$\alpha_G$, $M_G$, $\epsilon_3$  & 3 \\
SUSY (GUT scale)  & $m_{16}$, $M_{1/2}$, $A_0$, $m_{H_u}$, $m_{H_d}$  & 5  &
$m_{16}$, $M_{1/2}$, $\a$, $A_0$, $m_{H_u}$, $m_{H_d}$  & 6 \\
textures          & $\lambda$                                         & 1  &
$\lambda$  & 1 \\
SUSY (EW scale)   & $\tan \beta$, $\mu$                               & 2  &
$\tan \beta$, $\mu$    & 2 \\
\hline
Total \#          &                                                   & 11 &
                                                                            &
12\\
\hline
\end{tabular}
}
\caption{\footnotesize The $SO(10)$ GUT models with universal and non-universal
gaugino masses are both defined by three gauge parameters, $\alpha_{G}, M_{G},
\epsilon_3$; one large Yukawa coupling, $\lambda$; $\mu,$ and $\tan\beta$ are
obtained at the weak scale by consistent electroweak symmetry breaking. There
are 5 SUSY parameters defined at the GUT scale: $m_{16}$ (universal scalar mass
for squarks and sleptons), $M_{1/2}$ (universal gaugino mass), $A_0$ (universal
trilinear scalar coupling), and $m_{H_u}, \ m_{H_d}$ (up and down Higgs masses).
The models with non-universal gaugino masses have one additional parameter in
the SUSY sector, $\a$, which is the ratio of anomaly mediation to gravity mediation
contribution to gaugino masses.}
\label{tab:parameters}
\end{center}
\end{table}

\subsection{Universal Gaugino Masses}
The minimal Yukawa-unified \SO{10} SUSY GUT is defined by universal squark and slepton masses, $m_{16}$, a universal cubic scalar parameter, $A_0$, universal gaugino masses, $M_{1/2}$, and non-universal Higgs masses, or ``just so'' Higgs splitting, $m_{H_u}$, $m_{H_d}$ or  $m^2_{H_{u(d)}}=m^2_{10}\left(1-(+)\Delta^2_{m_H}\right)$.\footnote{ Here, we consider ad hoc splitting between the up- and the down-type Higgs masses. In principle, there can be D-term contributions~\cite{Baer:2009ie} that can give rise to splitting between the Higgs masses of the form $m^2_{H_{u(d)}}=m^2_{10} -(+)2D \;$. Note that these D-term contributions also give additional splitting to the right- and left-handed scalars.} Radiative electroweak symmetry breaking requires $\Delta^2_{m_H} \approx 13\%$. Furthermore, $\mu$ is taken to be positive. Fitting the top, bottom, and tau masses restricts to the region of SUSY breaking parameter space with
\be
A_0 \approx -2m_{16},\; m_{10} \approx \sqrt{2}m_{16},\; m_{16} > \text{few
TeV},\; \mu,M_{1/2} \ll m_{16};
\ee
and, due to Yukawa unification of the third family at the GUT scale,
\be
\tanb \approx 50.
\ee
The $\chi^2$ analysis favors a large universal scalar mass, $m_{16} \simeq20$ TeV~\cite{Anandakrishnan:2012tj}. Benchmark points for the universal gaugino mass scenario are shown in~\ref{tab:benchmark-spectrum-uni}. The main features of the low-energy spectrum are as follows. The choice of the soft-terms at the GUT scale results in an inverted scalar mass hierarchy at the weak scale with the first and second generation scalars being much heavier than the third family. The first and second family squarks have masses equal to $m_{16}$ while the third generation squarks are much lighter. Fitting the branching ratio $\mathcal{B}(B_s \ra \mu^+ \mu^-)$ leads to a heavy CP-odd Higgs mass and places us in the decoupling limit. Thus the light Higgs is predicted to be Standard Model-like. The gauginos are much lighter than the scalars and thus the particles in this spectrum that are accessible at the LHC are the gluinos, the lightest neutralino, and the lightest chargino. Furthermore, as shown in~\ref{tab:benchmark-spectrum-uni}, the $\chi^2$ analysis favors lighter gluino masses and the model will be in tension with low energy observables (especially
the 125 GeV Higgs mass) if the gluinos are not discovered to be lighter than $\simeq 2 $ TeV~\cite{Anandakrishnan:2012tj}. The gluino branching ratios of the benchmark points for the universal case are calculated using \code{SDECAY}~\cite{Muhlleitner:2003vg} and given in~\ref{tab:benchmark-decays-uni}. The dominant branching ratio of the gluino is $\mathcal{B}(\widetilde{g}\ra t b \widetilde{\chi}^{\pm}_1)$. The remaining contributions to the gluino decays are to $t \bar{t}\widetilde{\chi}^0_i$ and $b \bar{b} \widetilde{\chi}^0_i$ (i=1,2). $\widetilde{\chi}^0_2$ is wino-like and relatively light so it plays a role in gluino decays. The lightest neutralino is bino-like and, if the dominant dark matter candidate, would over-close the universe. 

\begin{table}[h!]
\centering
\renewcommand{\arraystretch}{1.2}
\begin{tabular}{|l|r|r|r|r|r|r|r|}
\hline
Universal & Ua & Ub & Uc  & Ud &  Ue & Uf \\ 
\hline
$M_{1/2}$ & 150 & 200 & 250  & 300 &  400 & 600 \\ 
$\mu$ & 869 & 890 & 824 & 879 & 924 & 974 \\ 
\hline
$\chi^2/$d.o.f & 0.11 & 0.21 & 0.38 & 0.58 &  1.12 & 1.98 \\ 
\hline
$M_A$ & 2300 & 2323 & 2263 & 2300 & 2361 & 2751 \\
$M_{\widetilde{g}}$ & 802 & 932 & 1061 & 1187 & 1309 & 1430 \\ 
$M_{\widetilde{t}_1}$ & 3776 & 3790 & 3696 & 3728 & 3760 & 3776 \\ 
$M_{\widetilde{b}_1}$ & 4634 & 4654 & 4580 & 4608 & 4640 & 4628 \\ 
$M_{\widetilde{\tau}_1}$ & 7865 & 7886 & 7834 & 7861 & 7896 & 7890 \\ 
$M_{\widetilde{\chi}^0_1}$ & 129 & 151 & 173 & 195 & 217 & 239 \\ 
$M_{\widetilde{\chi}^0_2}$ & 264 & 303 & 342 & 382 & 422 & 461 \\ 
$M_{\widetilde{\chi}^0_3}$ & 873 & 893 & 828 & 882 & 927 & 977 \\
$M_{\widetilde{\chi}^0_4}$ & 876 & 897 & 833 & 887 & 932 & 982 \\ 
$M_{\widetilde{\chi}^+_1}$ & 264 & 303 & 342 & 382 & 422 & 461 \\
$M_{\widetilde{\chi}^+_2}$ & 877 & 898 & 833 & 888 & 933 & 982 \\ 
\hline
\end{tabular}
\caption{Benchmark points of the universal case with $m_{16} = 20$ TeV. The benchmark points have been chosen with varying values of gluino mass. The model provides good fits to all low-energy observables at 95\% C.L.s for $M_{\widetilde{g}} < 2000$ GeV~\cite{Anandakrishnan:2012tj}. The above fits were obtained by fixing $m_{16}$ and $M_{1/2}$ (to obtain unique gluino masses) and thus yielding 2 d.o.f. for this system.}
\label{tab:benchmark-spectrum-uni}
\end{table}
\begin{table}[h!]
\centering
\renewcommand{\arraystretch}{1.5}
\begin{tabular}{|l|r|r|r|r|r|r|r|}
\hline
Universal & Ua & Ub & Uc  & Ud &  Ue & Uf \\ 
\hline
$\mathcal{B}(\widetilde{g}\ra g \widetilde{\chi}^0_i)$ & 0\% & 0\% & 2\% & 2\% &
2\% & 2\% \\ 
$\mathcal{B}(\widetilde{g}\ra t \bar{t} \widetilde{\chi}^0_1)$ & 6\% & 7\% & 7\%
& 7\% & 8\% & 8\% \\
$\mathcal{B}(\widetilde{g}\ra t \bar{t} \widetilde{\chi}^0_2)$ & 8\% & 11\% &
13\% & 14\% & 15\% & 15\% \\ 
$\mathcal{B}(\widetilde{g}\ra b \bar{b} \widetilde{\chi}^0_1)$ & 5\% & 4\% & 3\%
& 3\% & 3\% & 3\% \\ 
$\mathcal{B}(\widetilde{g}\ra b \bar{b} \widetilde{\chi}^0_2)$ & 18\% & 16\% &
14\% & 13\% & 12\% & 11\% \\ 
$\mathcal{B}(\widetilde{g}\ra t b \widetilde{\chi}^{\pm}_1)$ & 62\% & 62\% &
60\% & 60\% & 58\% & 56\% \\ 
$\mathcal{B}(\widetilde{g}\ra t b \widetilde{\chi}^{\pm}_2)$ & 0\% & 0\% & 0\% &
0\% & 2\% & 2\% \\ 
\hline
\end{tabular}
\caption{ Gluino decay branching ratios into different final states
for the six benchmark models 
given in~\ref{tab:benchmark-spectrum-uni}. For each model, we give the dominant
branching fractions. The loop decays 
in the first row include all four neutralinos. These ratios were calculated
using \code{SDECAY}~\cite{Muhlleitner:2003vg}.}
\label{tab:benchmark-decays-uni}
\end{table}

\subsection{Non-universal gaugino masses}
Non-universal gaugino masses can be obtained by considering a hybrid SUSY breaking mechanism. In Ref.~\cite{Anandakrishnan:2013cwa}, it was shown that Yukawa unification could be obtained with gravity and anomaly mediation contributions to the gaugino masses~\cite{Choi:2005ge,Choi:2005hd,Kitano:2005wc,Lowen:2008fm}. The gaugino masses are given by

\be
M_i = \left( 1 + \frac{g^2_G b_i
\a}{16\pi^2}\text{log}\left(\frac{M_{Pl}}{m_{16}}\right) \right) M_{1/2} \; ,
\ee

where $b_i=(33/5,1,-3)$ for $i=1,2,3$, $M_{1/2}$ is the overall mass scale, and $\a$ is the ratio of the anomaly mediation to gravity mediation contributions. The size of $\a$ plays a crucial role in determining the ratio of the gaugino masses and in addition the spectrum that is consistent with Yukawa unification. 
When $\a > 4$, the gaugino masses $M_1, M_2$ and $M_3$ have opposite signs. We refer to this region of parameter space as ``large $\a$''. For $\a < 4$, or ``small $\a$'', the $M_i$ all have the same sign. The relative sign of the gaugino masses leads to qualitatively different low energy spectra. We elaborate below on the benchmark models obtained for various values of $\a$. Note that values of $\a$ between 3 and 4 cause the gluino to be the LSP. We do not consider this region. The other soft-terms are similar to the case of universal gaugino masses. There is a universal scalar mass for the squarks and sleptons, $m_{16}$, a universal cubic scalar parameter, $A_0$, and non-universal Higgs mass parameters, $m_{H_u}$ and $m_{H_d}$. Finally, $\tanb \approx 50$.

\subsubsection{Small \texorpdfstring{$\bs{\a}$}{pdf} } 

\begin{table}[h!]
\centering
\renewcommand{\arraystretch}{1.2}
\begin{tabular}{|l|r|r|r|r|}
\hline
Small $\a$ & DMa & DMb & COa  & COb \\
\hline
$\epsilon_3$ & 0.00&0.00 &0.01 & 0.01 \\
$M_{1/2}$ & 450  & 600 & 450 & 485 \\
$\a$ & 1.5 &2.3 &2.54 & 2.61 \\
$\mu$ & 660 & 1199 & 1027 & 1035 \\
$m_{16}$ & 20000 & 29781 & 20000& 20000 \\
\hline
$\chi^2/$d.o.f & 0.92 & 0.86 & 1.07  & 1.19 \\
\hline
$M_A$ & 1915 & 3093 & 2442 & 3069\\ 
$M_{\widetilde{g}}$ &  1130 & 1135 & 707 &697\\
$M_{\widetilde{t}_1}$ & 3612 & 5832 & 3928 & 3951\\
$M_{\widetilde{b}_1}$ &  4770 & 7543 & 5453 & 5431\\
$M_{\widetilde{\tau}_1}$ & 6867 & 10565 & 8026& 8024 \\
$M_{\widetilde{\chi}^0_1}$ & 474 & 799 & 614 & 655\\
$M_{\widetilde{\chi}^0_2}$ & 557 & 836 & 629 & 683\\
$M_{\widetilde{\chi}^0_3}$ &  663 & 1201 & 1030& 1038\\
$M_{\widetilde{\chi}^0_4}$ &  694 & 1211 & 1037& 1047\\
$M_{\widetilde{\chi}^+_1}$ &  555 & 836 & 615& 656\\
$M_{\widetilde{\chi}^+_2}$ & 691 & 1210 &1037  & 1045\\
\hline
$\Omega h^2$ & 0.121 & 0.099 & 0.011 & 0.011\\
\hline
\end{tabular}
\caption{Benchmark points for models consistent with Yukawa unification with small non-universal contributions to the gaugino masses. The gaugino spectrum becomes increasingly compressed with increasing $\a$. The well-tempered dark matter points were obtained by fixing $\mu$, $M_{1/2}$ and $\a$ to get the right admixture for the LSP. The compressed spectrum points were obtained by fixing $m_{16}$, $M_{1/2}$, and $\a$ to obtain the small splitting between the gluino and the neutralino. There were 2 d.o.f. in determining the best fit points shown above.}
\label{tab:benchmark-spectrum-sma}
\end{table}
\begin{table}[h!]
\centering
\renewcommand{\arraystretch}{1.2}
\begin{tabular}{|l|r|r|r|r|}
\hline
Small $\a$ & DMa & DMb & COa  & COb \\
\hline
$BR(\widetilde{g}\ra g \widetilde{\chi}^0_1)$ & 0\% & 2\% & 41\% & 83\% \\
$BR(\widetilde{g}\ra g \widetilde{\chi}^0_2)$ & 1\% & 2\% & 3\% & 1\% \\
$BR(\widetilde{g}\ra g \widetilde{\chi}^0_3)$ & 6\% & 0\% & 0\% & 0\% \\
$BR(\widetilde{g}\ra g \widetilde{\chi}^0_4)$ & 4\% & 0\% & 0\% & 0\% \\
$BR(\widetilde{g}\ra t \bar{t} \widetilde{\chi}^0_1)$ & 2\% & 0\% & 0\% & 0\% \\
$BR(\widetilde{g}\ra t \bar{t} \widetilde{\chi}^0_2)$ & 4\% & 0\% & 0\% & 0\% \\
$BR(\widetilde{g}\ra t \bar{t} \widetilde{\chi}^0_3)$ & 4\% & 0\% & 0\% & 0\% \\
$BR(\widetilde{g}\ra t \bar{t} \widetilde{\chi}^0_4)$ & 3\% & 0\% & 0\% & 0\% \\
$BR(\widetilde{g}\ra b \bar{b} \widetilde{\chi}^0_1)$ & 4\% & 14\% & 51\% & 15\% \\
$BR(\widetilde{g}\ra b \bar{b} \widetilde{\chi}^0_2)$ & 9\% & 38\% & 2\% & 0\% \\
$BR(\widetilde{g}\ra b \bar{b} \widetilde{\chi}^0_3)$ & 7\% & 0\% & 0\% & 0\% \\
$BR(\widetilde{g}\ra b \bar{b} \widetilde{\chi}^0_4)$ & 4\% & 0\% & 0\% & 0\% \\
$BR(\widetilde{g}\ra t b \widetilde{\chi}^{\pm}_1)$ & 32\% & 42\% & 0\% & 0\% \\
$BR(\widetilde{g}\ra t b \widetilde{\chi}^{\pm}_2)$ & 20\% & 0\% & 0\% & 0\% \\
\hline
\end{tabular}
\caption{ Gluino decay branching ratios into different final states for the DM (dark matter) and CO (compressed gaugino spectrum) benchmark models given in~\ref{tab:benchmark-spectrum-sma}. For each model, we give the dominant branching fractions. These ratios were calculated using \code{SDECAY}~\cite{Muhlleitner:2003vg}.}
\label{tab:benchmark-decays-sma}
\end{table}

When $\a < 4$, there are small non-universal contributions to the gaugino masses. The additional degree of freedom, $\a$, allows tuning of the ratios of $M_1$ and $M_2$. As $\alpha$ is gradually increased from $0$, there are two effects to the spectrum. The first one is that the wino component of the LSP begins to increase. Therefore, it is possible to obtain the correct relic density by tuning $M_{1/2}$, $\a$, and $\mu$. Secondly, since the beta-function coefficient is negative for SU(3), the gluino mass decreases with increasing $\a$ until it becomes the lightest supersymmetric particle for $\a \gtrsim 3$. The region in the parameter space that fits the measured value of dark matter relic abundance is of special interest and was studied in Ref.~\cite{Anandakrishnan:2013tqa} (see also~\cite{Lowen:2008fm,Choi:2005uz,Krippendorf:2013dqa}). We show two sample benchmark models which reproduce the measured relic abundance~\cite{Ade:2013ktc} in~\ref{tab:benchmark-spectrum-sma}. Point DMa is well-tempered bino-wino-higgsino mixture whereas point DMb has a bino-wino mixture.  The spin independent scattering cross-section for the benchmark point DMa is $1.6 \times 10^{-8}$ pb. Note that this benchmark point is now ruled out by LUX results~\cite{Akerib:2013tjd}, which exclude spin independent scattering cross-sections down to $6 \times 10^{-9}$ pb. An LSP of higher mass can be obtained by increasing the values of $M_{1/2}$ and $\mu$ while maintaining the relic density with bino-wino-higgsino well-tempering and be consistent with the LUX result. On the other hand, the spin independent scattering cross section for point DMb is below the bounds from LUX at $3.5 \times 10^{-9}$ pb.  The dominant decay modes of the gluino for points DMa and DMb are also very similar to the universal case with large decay fractions into the 3-body modes: $t b \widetilde{\chi}^\pm $, $ t \bar{t} \widetilde{\chi}^0$, $b \bar{b} \widetilde{\chi}^0$.

Another region of interest is when $\a$ is slightly less than 3 and the gauginos are nearly degenerate~\cite{Lowen:2008fm,Choi:2005uz}. In this region, the gluino is only a few GeV heavier than the LSP. Sample benchmark points, COa and COb, are presented in~\ref{tab:benchmark-spectrum-sma}. The phenomenology of this region is interesting since the decay products of the gluino will be soft and may not be visible at the LHC. Also, as shown in~\ref{tab:benchmark-decays-sma}, the loop decays of the gluino into $g \widetilde{\chi}^0$ start dominating over the 3-body decays. The LSP contains a large wino component and the relic abundance is not saturated for these values of $\a$. 

The best fit points obtained with $\a < 4$ are very similar to the benchmark points discussed in the case of universal gaugino masses. The first two family of scalars are very heavy ($\sim m_{16}$), and the spectrum has the inverted scalar mass hierarchy. The extra MSSM Higgses are heavy and decoupled and thus the lightest Higgs is purely SM like. The benchmark points prefer the value of $\epsilon_3 \approx 0-1\%$. $\epsilon_3$ is the GUT scale threshold correction required to fit the value of $\alpha_s(M_Z)$. While standard MSSM scenarios with universal gaugino mass prefer $\epsilon_3 = -3\%$, it was shown in Ref.~\cite{Krippendorf:2013dqa,Raby:2009sf} that precision gauge coupling unification can be achieved in models with non-universal gaugino masses, especially with the mirage pattern with lighter gluinos considered in this work. 

\subsubsection{Large \texorpdfstring{$\bs{\a}$}{pdf}} 

For $\a > 4$, $M_1$ and $M_2$ have the opposite sign as $M_3$. We choose $\mu,M_{1/2}<0$ and thus fix $M_3>0,\;M_1,M_2<0$. The first and second family squarks and sleptons have mass of order $m_{16} = 5$ TeV while the third family scalars are roughly a factor of 2 lighter.  In addition, gluinos are always lighter than the third family squarks and sleptons, and the lightest charginos and neutralinos are even lighter. The third generation squarks have mass $\sim$1.6 TeV and the lightest neutralino and the lightest chargino are nearly degenerate. The LSP is wino-like and, if the dominant thermal dark matter component, produces an under-abundant dark matter relic. Since the chargino and neutralino are nearly degenerate, 1-loop corrections to the mass difference has been estimated based on Ref.~\cite{Cheng:1998hc}. If the mass difference is less than $\simeq$ 1.5 GeV, then the charginos decay to the neutralino and pions. These decay widths have also been calculated using Ref.~\cite{Chen:1996ap}.
   
The particles in the spectrum that are accessible at the LHC are the gluinos, the two lightest neutralinos, and the lightest chargino. In contrast to the universal case, the $\chi^2$ analysis does not favor any specific range of gluino masses (cf.~\ref{tab:benchmark-spectrum-mir}). The masses and gluino branching ratios of the benchmark points for the large $\a$ case are given in~\ref{tab:benchmark-spectrum-mir} and~\ref{tab:benchmark-decays-mir}, respectively. The dominant branching ratio of the gluino is $\mathcal{B}(\widetilde{g}\ra t b \widetilde{\chi}^{\pm}_i)$ (i=1,2). The remaining contributions to the gluino decays are to $t \bar{t} \widetilde{\chi}^0_j$ and $b \bar{b} \widetilde{\chi}^0_j$ (j=1,2,3). The neutralinos are more degenerate so the heavier ones contribute a non-negligible amount to the branching ratios.
\begin{table}[h!]
\centering
\renewcommand{\arraystretch}{1.2}
\begin{tabular}{|l|r|r|r|r|r|r|}
\hline
Large $\a$ & Ma & Mb & Mc  & Md &  Me & Mf \\
\hline
$M_{1/2}$ & -181 & -200 & -220 & -200 & -200 & -200 \\
$\mu$ & -341 & -341 & -346 & -355 & -376 & -405  \\
$\a$ & 6 & 8 & 8 & 9 & 10 & 11\\
\hline
$\chi^2$/d.o.f. & 0.36 & 0.40 & 0.45 & 0.41 & 0.41 & 0.42 \\
\hline
$M_A$ & 2549 & 2513 & 2490 & 2510 & 2453 & 2383  \\ 
$M_{\widetilde{g}}$ & 859 & 940 & 1025 & 1107 & 1270 & 1429  \\
$M_{\widetilde{t}_1}$ & 1717 & 1734 & 1764 & 1799 & 1864 & 1937  \\
$M_{\widetilde{b}_1}$ & 1490 & 1522 & 1561 & 1592 & 1680 & 1780 \\
$M_{\widetilde{\tau}_1}$ & 2132 & 2124 & 2126 & 2128 & 2131 & 2148 \\
$M_{\widetilde{\chi}^0_1}$ & 276 & 290 & 305 & 307 & 327 & 352  \\
$M_{\widetilde{\chi}^0_2}$ & 347 & 346 & 351 & 360 & 381 & 409 \\
$M_{\widetilde{\chi}^0_3}$ & 389 & 405 & 428 & 423 & 443 & 467   \\
$M_{\widetilde{\chi}^0_4}$ & 576 & 632 & 693 & 698 & 764 & 830  \\
$M_{\widetilde{\chi}^+_1}$ & 279 & 294 & 309 & 310 & 331 & 355 \\
$M_{\widetilde{\chi}^+_2}$ & 395 & 409 & 431 & 426 & 447 & 470  \\
\hline
\end{tabular}
\caption{Benchmark points of the mirage case with $m_{16} = 5$ TeV. The gaugino sector of the model is different from the universal case, good fits are obtained for heavier gluino masses and the lightest chargino and neutralino are nearly degenerate. In addition to $m_{16}$, $M_{1/2}$ and $\a$ were also held fixed in order to obtain unique values of the gluino mass.}
\label{tab:benchmark-spectrum-mir}
\end{table}

\begin{table}[h!]
\centering
\renewcommand{\arraystretch}{1.2}
\begin{tabular}{|l|r|r|r|r|r|r|r|}
\hline
Large $\a$ & Ma & Mb & Mc  & Md &  Me & Mf \\
\hline
$\mathcal{B}(\widetilde{g}\ra g \widetilde{\chi}^0_1)$ & 0\% & 0\% & 0\% & 0\% &
0\% & 0\%  \\
$\mathcal{B}(\widetilde{g}\ra g \widetilde{\chi}^0_2)$ & 2\% & 1\% & 1\% & 1\% &
1\% & 0\%   \\
$\mathcal{B}(\widetilde{g}\ra g \widetilde{\chi}^0_3)$ & 1\% & 0\% & 0\% & 0\% &
0\% & 0\%  \\
$\mathcal{B}(\widetilde{g}\ra g \widetilde{\chi}^0_4)$ & 0\% & 0\% & 0\% & 0\% &
0\% & 0\%   \\
$\mathcal{B}(\widetilde{g}\ra t \bar{t} \widetilde{\chi}^0_1)$ & 7\% & 8\% & 9\%
& 8\% & 9\% & 9\%  \\
$\mathcal{B}(\widetilde{g}\ra t \bar{t} \widetilde{\chi}^0_2)$ & 4\% & 6\% & 7\%
& 8\% & 8\% & 9\%  \\
$\mathcal{B}(\widetilde{g}\ra t \bar{t} \widetilde{\chi}^0_3)$ & 1\% & 1\% & 1\%
& 1\% & 2\% & 3\%  \\
$\mathcal{B}(\widetilde{g}\ra t \bar{t} \widetilde{\chi}^0_4)$ & 0\% & 0\% & 0\%
& 0\% & 0\% & 0\%   \\
$\mathcal{B}(\widetilde{g}\ra b \bar{b} \widetilde{\chi}^0_1)$ & 19\% & 16\% &
15\% & 14\% & 12\% & 11\%   \\
$\mathcal{B}(\widetilde{g}\ra b \bar{b} \widetilde{\chi}^0_2)$ & 13\% & 13\% &
12\% & 12\% & 11\% & 11\%   \\
$\mathcal{B}(\widetilde{g}\ra b \bar{b} \widetilde{\chi}^0_3)$ & 7\% & 6\% & 6\%
& 6\% & 6\% & 6\%   \\
$\mathcal{B}(\widetilde{g}\ra b \bar{b} \widetilde{\chi}^0_4)$ & 0\% & 0\% & 0\%
& 0\% & 0\% & 0\%   \\
$\mathcal{B}(\widetilde{g}\ra t b \widetilde{\chi}^{\pm}_1)$ & 30\% & 32\% &
36\% & 32\% & 32\% & 32\%  \\
$\mathcal{B}(\widetilde{g}\ra t b \widetilde{\chi}^{\pm}_2)$ & 16\% & 14\% &
12\% & 16\% & 18\% & 20\%  \\
\hline
\end{tabular}
\caption{Gluino decay branching ratios into different final states for the eight benchmark models given in~\ref{tab:benchmark-spectrum-mir}. For each model, we give the dominant
branching fractions. These ratios were calculated using \code{SDECAY}~\cite{Muhlleitner:2003vg}.}
\label{tab:benchmark-decays-mir}
\end{table}

\section{Experimental Analyses and Results}

The searches for new physics by the ATLAS and CMS collaborations are often interpreted within the context of simplified models. Such models are intended to capture the main features of specific scenarios with the risk being that the assumptions are over-simplified. The simplified models studied by the collaborations that most resemble our models assume branching ratios of $\mathcal{B}(\widetilde{g}\rightarrow t\bar{t}\widetilde{\chi}^0_1)=100\%$, 
$\mathcal{B}(\widetilde{g}\rightarrow b\bar{b}\widetilde{\chi}^0_1)=100\%$, or $\mathcal{B}(\widetilde{g}\rightarrow tb\widetilde{\chi}^{\pm}_1)=100\%$. However, the branching ratios for our models clearly do not match the decay branching ratio for any one simplified model. Fortunately, the ATLAS and CMS collaborations provide, for a given search and signal region therein, an allowed number of events from new physics. It is thus possible to reinterpret the searches for new physics in the context of more elaborate models. The contribution of a sparticle to the number of events in a signal region is a function of the production cross section of the sparticle and the sparticle's branching fraction into final states relevant to the signal region. By comparing this contribution to the allowed number of events from new physics, an exclusion limit can be set on the sparticle's mass. For both boundary conditions considered in this paper, the squarks and sleptons are sufficiently heavier than the gauginos and thus do not contribute to the final states relevant for gluino or electroweakino\footnote{Electroweakinos refer to neutralinos and charginos, collectively.} searches. Furthermore, searches designed to look for gluinos do not overlap with those designed to look for electroweakinos. That is, electroweakinos do not contribute to signal regions from searches for gluinos, and vice versa. This is at least true for our models. Since there is no contamination from other sparticles, we are able to directly place bounds on the gluino masses in our models.

The procedure of Ref.~\cite{Anandakrishnan:2013nca} was performed for each benchmark model considered here to determine the limits from the 3 CMS analyses. In addition, the program \code{CheckMATE}\footnote{\code{CheckMATE} uses~\code{Delphes 3}~\cite{deFavereau:2013fsa},~\code{FastJet}~\cite{Cacciari:2011ma,Cacciari:2005hq}, and the Anti-kt jet algorithm~\cite{Cacciari:2008gp}.} was used to evaluate bounds on the gluino mass for each model. \code{CheckMATE} requires as input a \code{HepMC}~\cite{Dobbs:2001ck} file containing generated events and the production cross section of the sparticles of interest along with the total $1\s$ uncertainty on the cross section. We use \code{PYTHIA 8.175}~\cite{Sjostrand:2007gs} to generate 10,000--20,000 events in \code{HepMC} format.\footnote{We are using \code{HepMC 2.06.09}.} The gluino production cross section and its uncertainty are obtained from~\cite{Kramer:2012bx}. It was found that the ATLAS analysis ATLAS-CONF-2013-061 ~\cite{ATLAS-CONF-2013-061}, on which we elaborate below, is the most constraining analysis for each of the benchmark points in our models with the exception of COa and COb. These two points were not ruled out by any analysis considered in this paper. The analysis which came the closest to ruling out these points was the ATLAS analysis ATLAS-CONF-2013-047~\cite{ATLAS-CONF-2013-047}, which is a search for final states with high-$\pt$ jets, missing transverse momentum, and no electrons or muons. Both of these analyses are implemented in \code{CheckMATE}.

The ATLAS-CONF-2013-061 analysis is a search for final states with large missing transverse momentum, at least four, six, or seven jets, at least three jets tagged as $b$-jets, and either zero or at least one lepton. It was performed at $\sqrt{s}=8$ TeV with 20.1 fb$^{-1}$ of data. The results are interpreted in the context of a variety of simplified models. The simplified models considered that are relevant to our models are the Gbb, Gtt, and Gtb models. These models assume 100\% branching ratios of a gluino to $b \bar{b}\widetilde{\chi}^0_1$, $t \bar{t} \widetilde{\chi}^0_1$, and $t b\widetilde{\chi}^{\pm}_1$, respectively. The most constraining signal regions on our models are presented in~\ref{tab:signal-region-1lep} for the universal and dark matter cases and in \ref{tab:signal-region-0lep} for the large $\a$ case. The universal and dark matter cases are constrained most by the signal region requiring at least 1 lepton and at least 6 jets while the large $\a$ case is most constrained by the signal region requiring no leptons and at least 4 jets. Each of the benchmark points in these cases has a sizeable gluino branching fraction into $t b \widetilde{\chi}^{\pm}_1$. For the universal and dark matter points, the mass splitting of the lightest chargino and the lightest neutralino is large enough for the chargino to decay to a $W$ whose decay products are energetic enough to be seen in the detector. This is not true for the large $\a$ case. The lightest chargino and the lightest neutralino in the large $\a$ benchmark points are nearly degenerate and so the charginos produced from gluino decays in these points are effectively missing energy.

\vspace{1cm}
\begin{table}[h!]
\centering
\renewcommand{\arraystretch}{1.5}
\begin{tabular}{|c|c|c|c|c|c|}
\hline
\mc{6}{|l|}{baseline selection: $\ge$ 1 signal lepton ($e,\mu$),
$p_\text{T}^{j_1} >$ 90 GeV, $\etmiss >$ 150 GeV,}\\
\mc{6}{|l|}{\hspace{3.17cm}$\ge$ 4 jets with $p_\text{T} >$ 30 GeV, $\ge 3$
$b$-jets with $p_\text{T} >$ 30 GeV}\\
\hline
Signal Region &  $N$ jets &  $\etmiss$ [GeV] & $m_\text{T}$ [GeV] &
$m^\text{incl}_\text{eff}$ [GeV] 
& $\etmiss$/ $\sqrt{H^\text{incl}_\text{T}}$ [GeV$^\half$]\\
\hline
SR-1l-6J-B & $\ge$ 6 & $>$ 225 & $>$ 140  & $>$ 800 & $>$ 5 \\
\hline
\end{tabular}
\caption{ Most constraining signal region for the universal and dark matter scenarios. A detailed description of the parameters in this table can be found in~\cite{ATLAS-CONF-2013-061}.}
\label{tab:signal-region-1lep}
\end{table}

\begin{table}[h!]
\centering
\renewcommand{\arraystretch}{1.5}
\begin{tabular}{|c|c|c|c|c|}
\hline
\mc{5}{|l|}{baseline selection: baseline lepton veto, $p_\text{T}^{j_1} >$ 90
GeV, $\etmiss >$ 150 GeV, $\ge$ 4 jets with $p_\text{T} >$ 30 GeV,}\\
\mc{5}{|l|}{\hspace{3.1cm} $\Delta\phi^{4j}_\text{min} \ge$ 0.5,
$\etmiss$/$m^{4j}_\text{eff} >$ 0.2, $\ge$ 3 $b$-jets with $p_\text{T} >$ 30
GeV}\\
\hline
Signal Region &  $N$ jets &  $\etmiss$ [GeV] & $p_\text{T}$ jets [GeV] & $m^\text{4j}_\text{eff}$ [GeV] \\
\hline
SR-0l-4J-C & $\ge$ 4 & $>$ 250 & $>$ 50  & $>$ 1300 \\
\hline
\end{tabular}
\caption{ Most constraining signal region for the large $\a$ scenario. A detailed description of the parameters in this table can be found 
in~\cite{ATLAS-CONF-2013-061}.}
\label{tab:signal-region-0lep}
\end{table}
\vspace{1cm}
In the ATLAS analysis, the number of observed events in a given signal region is used to calculate a 95\% confidence level (CL) upper limit on the allowed number of events by new physics. Each of the simplified models (Gbb,Gtt,Gtb) has two free parameters: the gluino mass and the neutralino mass.\footnote{In the Gtb model, the mass difference between the chargino and neutralino is set to 2 GeV. So, there are still only two free parameters in this model.} These two parameters are scanned over for a simplified model and a predicted number of events for the various signal regions is calculated. By comparing the predicted number to the 95\% CL upper limit, an observed limit is found in the gluino-neutralino masses parameter space with $\pm 1\s$ theoretical uncertainty. In \code{CheckMATE}, the number of signal events $S$ is determined for each signal region of a given analysis. The total $1\s$ uncertainty on this number $\Delta S$ is calculated from both the statistical uncertainty, given by Monte Carlo generated events, and the systematic uncertainty, which is estimated from the total uncertainty on the signal cross section provided by the user. In order to compare results from \code{CheckMATE} to those in the ATLAS analysis, we calculate the ratio $r$ of the predicted number of signal events at the $1\s$ theoretical lower limit to the experimentally measured 95\% CL upper limit provided by the ATLAS analysis,\footnote{Note that by default \code{CheckMATE} calculates $r$ at the $2\s$ theoretical lower limit.}
\be
r \equiv \frac{S - \Delta S}{ S^{95}_\text{Exp} } \; .
\ee
A model is considered excluded if this ratio is $\ge$1. As a check, we calculate the bounds for the simplified models (Gbb,Gtt,Gtb) and obtain good agreement with the results of the ATLAS analysis (cf. Fig. 12 and Fig. 14 in~\cite{ATLAS-CONF-2013-061}).\footnote{We consider only individual signal regions when testing models and therefore compare our results to those in the ATLAS analysis coming from individual signal regions.} We then obtain exclusion limits on the gluino mass in our models. Points Ua-d are ruled out for the universal case and points Ma-d are ruled out for the large $\a$ case. In the dark matter scenario, the point DMa is ruled out while the point DMb is allowed. Point DMb is not well-constrained due to the small mass differences between the gluino and the lightest electroweakinos, $\Delta m_{\widetilde{g}-\widetilde{\chi}}\approx 300$~GeV (cf.~\ref{tab:benchmark-spectrum-sma}). 

\begin{figure}[h!]
\thisfloatpagestyle{empty}
\centering
\includegraphics[width=0.8\textwidth, trim=0ex 0ex 0ex 3ex, clip]{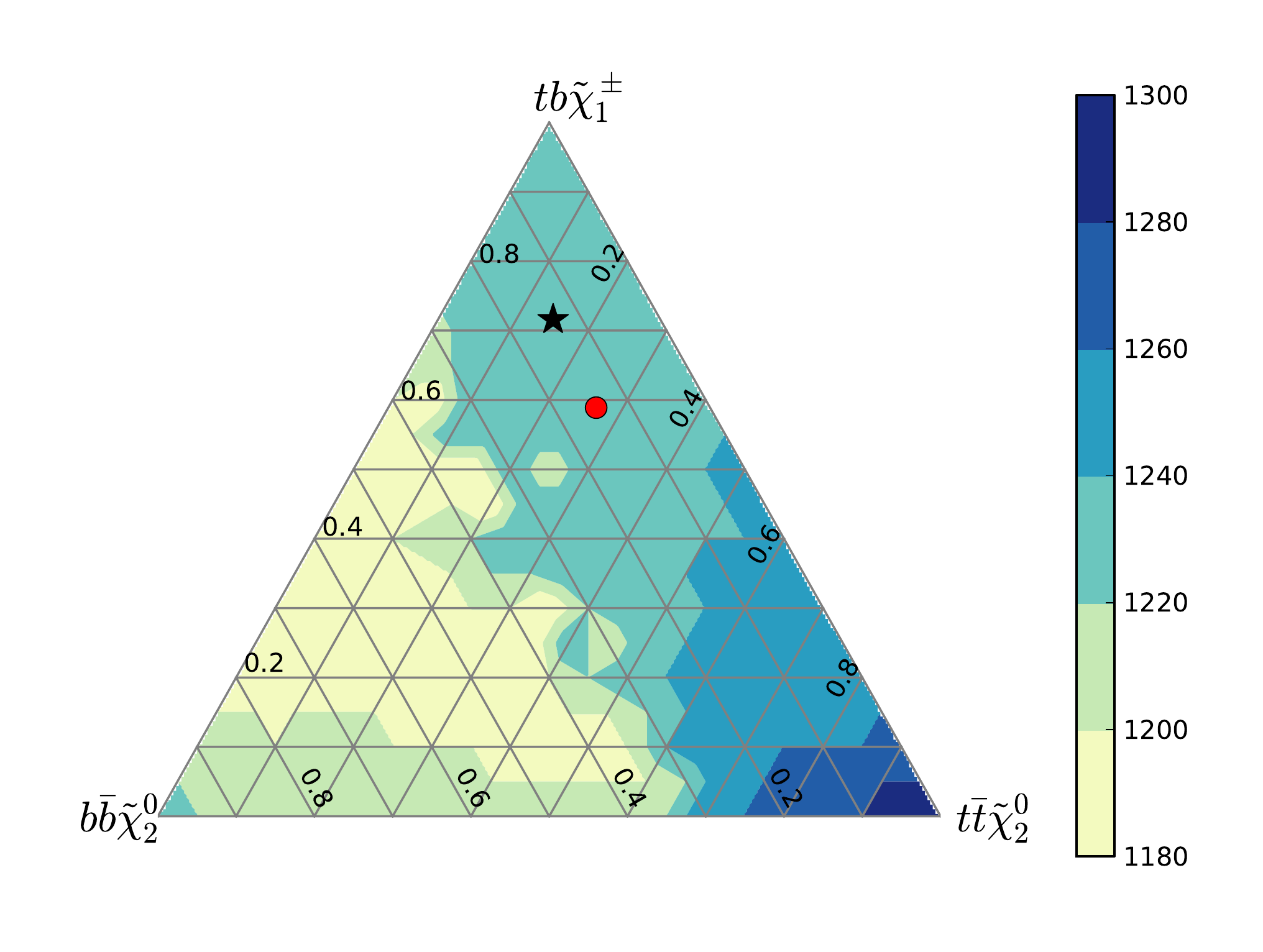}
\caption{\footnotesize Lowest gluino masses allowed by Ref.~\cite{ATLAS-CONF-2013-061} for models with branching fractions to combinations of $t b \widetilde{\chi}^{\pm}_1$, $b \bar{b} \widetilde{\chi}^0_2$, and $t \bar{t} \widetilde{\chi}^0_2$. All masses are in GeV. The black star and red dot are representative points for the universal and large $\a$ cases, respectively. We set $M_{\widetilde{\chi}^0_1}= 200$~GeV and $M_{\widetilde{\chi}^0_2}= 300$~GeV for the generation of these points.}
\label{fig:triangle}
\end{figure}

There are 6 dominant branching ratios in our models: $b \bar{b} \widetilde{\chi}^0_1$, $t \bar{t} \widetilde{\chi}^0_1$, $t b \widetilde{\chi}^{\pm}_1$, 
$b \bar{b} \widetilde{\chi}^0_2$, $t \bar{t} \widetilde{\chi}^0_2$, and $t b \widetilde{\chi}^{\pm}_2$. We find that the bounds for simplified models defined by 100\% branching fractions to $b \bar{b} \widetilde{\chi}^0_1$, $t \bar{t} \widetilde{\chi}^0_1$, or $t b \widetilde{\chi}^{\pm}_1$ are approximately equal. We also find that simplified models defined by 100\% branching fractions to $t \bar{t} \widetilde{\chi}^0_2$ and $t b \widetilde{\chi}^{\pm}_2$ receive nearly the same limits. We can therefore represent our models to a good approximation by models with branching fractions to combinations of $t b \widetilde{\chi}^{\pm}_1$, 
$b \bar{b} \widetilde{\chi}^0_2$, and $t \bar{t} \widetilde{\chi}^0_2$. As a complementary approach to testing the benchmark points directly and to obtain a more general representation of the bounds on models such as ours, we obtain exclusion limits for models defined by branching fractions to all combinations of these three decay modes and record the bounds on a triangle~\cite{Anandakrishnan:2014pva} as shown in~\ref{fig:triangle}. This approach is particularly useful for testing an approximately continuous range of gluino masses when the benchmark points have large differences between their gluino masses, as is the case for the universal and large $\a$ scenarios. While the range of exclusion limits shown in~\ref{fig:triangle} spans $\sim$100 GeV, it is clear that the ATLAS search ATLAS-CONF-2013-061 is sensitive to all six of the third-family-mediated gluino decay modes listed above.

In the universal case, each of the benchmark models shares nearly the same combination of branching ratios and can therefore be represented by a single point on the triangle. We represent this case by the black star in~\ref{fig:triangle}. This observation is also true for the large $\a$ case which we represent by a red dot. Note that the decay modes considered in the triangle only constitute $\sim$90\% of the branching ratios for models in the large $\a$ case. It is still useful however to see where these models lie in the triangle so that a comparison can be drawn with the universal case. These two cases inhabit the same region of the triangle and thus both  have exclusion limits on the gluino mass of $\sim$1.2 TeV. This is consistent with the limits obtained by testing the benchmark points directly.

\begin{figure}[h!]
\thisfloatpagestyle{empty}
\centering
\includegraphics[width=1.0\textwidth, trim=0ex 0ex 0ex 3ex, clip]{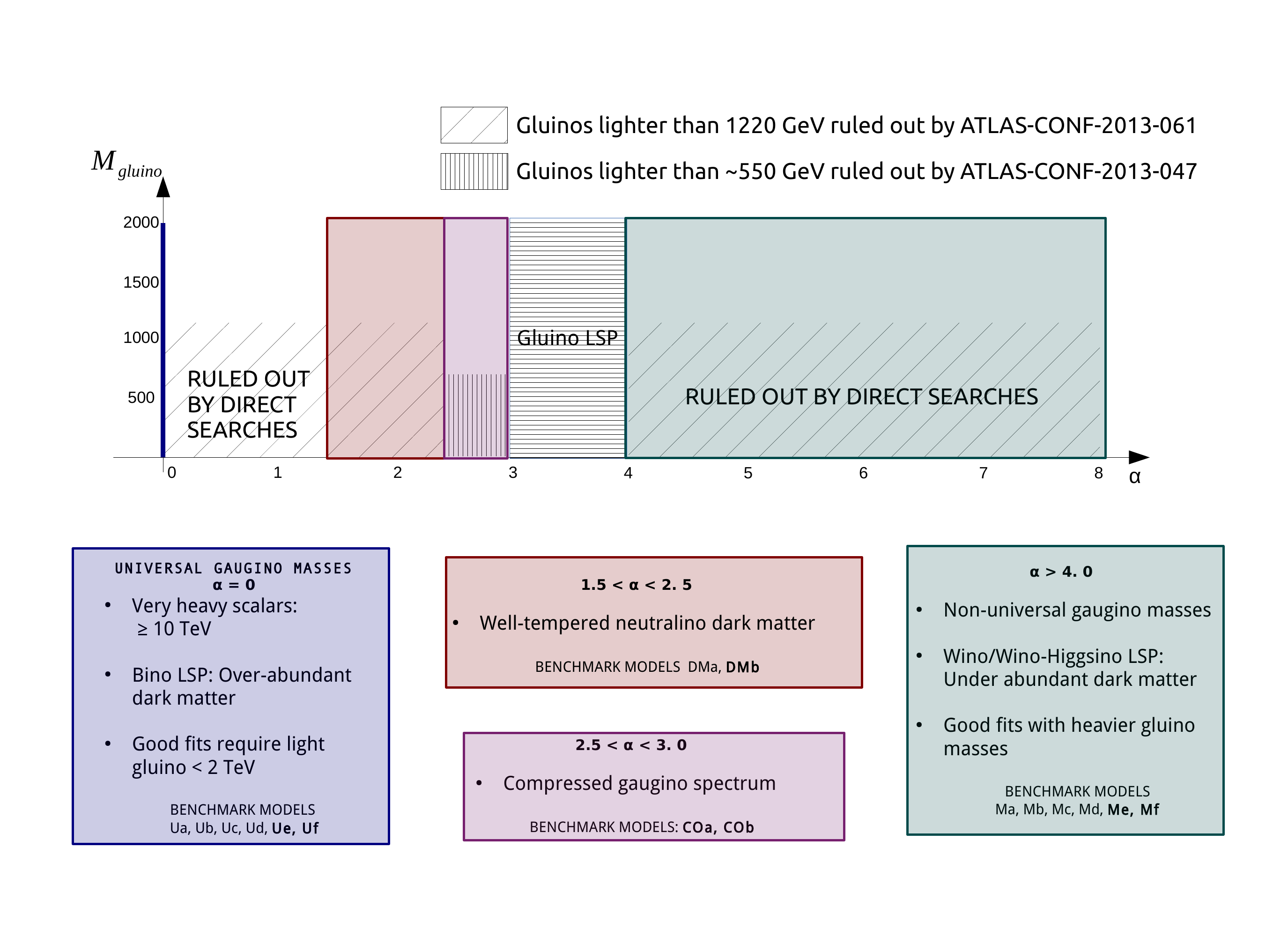}
\caption{\footnotesize Summary of the exclusion limits obtained for the benchmark models considered in this paper. The main features of each model scenario are listed in the boxes below the plot. Benchmark points listed in bold are not ruled out.}
\label{fig:summary}
\end{figure}

The exclusion limits on the mass of the gluino obtained for the benchmark models at the $1\s$ level are summarized in~\ref{fig:summary}. They are arranged with increasing $\a$ with the universal case corresponding to $\a=0$. Each model scenario in the plot is color coded with its main features listed in the box below the plot with corresponding color. Models in regions with diagonal slashes have been ruled out by the ATLAS search ATLAS-CONF-2013-061. Vertical black lines denote the region in which the benchmark points COa and COb have exclusion limits.\footnote{\label{footnote} The current bound on the gluino mass in such scenarios is $\sim$550 GeV~\cite{ATLAS-CONF-2013-047,Chatrchyan:2014lfa}.} The horizontal lines show the region in $\a$-space where models have a gluino LSP, which we did not consider in this paper. We find that the lower bound on $M_{\widetilde{g}}$ in Yukawa-unified SO(10) SUSY GUTs is generically $\sim$1.2 TeV at the $1\s$ level unless there is considerable degeneracy between the gluino and the LSP, in which case the bounds are much lower. 

\section{Conclusion}

In this paper, we surveyed the LHC phenomenology of Yukawa-unified SO(10) SUSY GUTs with various GUT scale boundary conditions. These boundary conditions are  characterized by (i) universal scalar masses and universal gaugino masses or by (ii) universal scalar masses and non-universal gaugino masses with effective 
mirage mediation. The size of the effective mirage mediation parameter, $\a$, plays a crucial role in determining the ratio of the gaugino masses and in addition the spectrum that is consistent with Yukawa unification. For boundary condition (ii), we considered three scenarios: (ii.1) roughly equal gravity mediated and anomaly mediated contributions to gaugino masses, (ii.2) well-tempered dark matter, and (ii.3) compressed gauginos. A global $\chi^2$ analysis was performed by varying GUT scale model parameters to fit low energy observables. Benchmark points were chosen based on the best fit of the Yukawa-unified SO(10) SUSY GUT models to low-energy data. The particles in the spectra of these models that are accessible at the LHC are the gluinos and the lightest electroweakinos. The focus of this paper was restricted to gluino phenomenology. The gluinos in models (i), (ii.1), and (ii.2) have large decay fractions into the 3-body modes, $t b \widetilde{\chi}^\pm $, $ t \bar{t} \widetilde{\chi}^0$, and $b \bar{b} \widetilde{\chi}^0$, while the branching fractions of the gluinos in model (ii.3) into $g \widetilde{\chi}^0$ are significant if not dominant.

Using recent experimental analyses relevant to gluino production at the LHC, we derived a lower bound on the gluino mass for each boundary condition except for (ii.3). Model (ii.3) was not constrained by any analysis considered. In this model, the gluino and the LSP are nearly degenerate so that the gluino decay products are too soft to be seen in the detector. We found that the lower bound on $M_{\widetilde{g}}$ in Yukawa-unified SO(10) SUSY GUTs is generically $\sim$1.2 TEV at the $1\s$  level unless there is considerable degeneracy between the gluino and the LSP, in which case the bounds are much lower (cf. footnote~\ref{footnote}). Hence many of our benchmark points (cf.~\ref{tab:benchmark-spectrum-uni},~\ref{tab:benchmark-spectrum-sma},~\ref{tab:benchmark-spectrum-mir}) are not ruled out by the present LHC data and are still viable models which can be tested at LHC 14. Finally, we note that current exclusion limits on simplified models in which the gluino decays to a pair of third family quarks and the lightest neutralino or chargino are $\sim$1.3 TeV at the $1\s$ level~\cite{ATLAS-CONF-2013-061,CMS-PAS-SUS-13-004}.

\section*{Acknowledgments}

We thank the authors of SDECAY for useful discussions. B.C.B.~and S.R.~received partial support for this work from DOE/ER/01545-902. A.A.~is supported on the Ohio State University Presidential Fellowship. We thank the \emph{Ohio Supercomputer Center} for their computing resources.


\bibliography{bibliography}

\providecommand{\href}[2]{#2}\begingroup\raggedright\begin{thebibliography}{10}

\bibitem{Dimopoulos:1981yj}
S.~Dimopoulos, S.~Raby, and F.~Wilczek, ``{Supersymmetry and the Scale of
  Unification},'' {\em Phys.Rev.} {\bf D24} (1981)
1681--1683.

\bibitem{Dimopoulos:1981zb}
S.~Dimopoulos and H.~Georgi, ``{Softly Broken Supersymmetry and SU(5)},'' {\em
  Nucl.Phys.} {\bf B193} (1981)
150.

\bibitem{Ibanez:1981yh}
L.~E. Ibanez and G.~G. Ross, ``{Low-Energy Predictions in Supersymmetric Grand
  Unified Theories},'' {\em Phys.Lett.} {\bf B105} (1981)
439.

\bibitem{Sakai:1981gr}
N.~Sakai, ``{Naturalness in Supersymmetric Guts},'' {\em Z.Phys.} {\bf C11}
  (1981)
153.

\bibitem{Einhorn:1981sx}
M.~Einhorn and D.~Jones, ``{The Weak Mixing Angle and Unification Mass in
  Supersymmetric SU(5)},'' {\em Nucl.Phys.} {\bf B196} (1982)
475.

\bibitem{Marciano:1981un}
W.~J. Marciano and G.~Senjanovic, ``{Predictions of Supersymmetric Grand
  Unified Theories},'' {\em Phys.Rev.} {\bf D25} (1982)
3092.

\bibitem{Blazek:2001sb}
T.~Blazek, R.~Dermisek, and S.~Raby, ``{Predictions for Higgs and supersymmetry
  spectra from SO(10) Yukawa unification with $\mu > 0$},'' {\em
  Phys.Rev.Lett.} {\bf 88} (2002) 111804,
\href{http://www.arXiv.org/abs/hep-ph/0107097}{{\tt hep-ph/0107097}}.

\bibitem{Blazek:2002ta}
T.~Blazek, R.~Dermisek, and S.~Raby, ``{Yukawa unification in SO(10)},'' {\em
  Phys.Rev.} {\bf D65} (2002) 115004,
\href{http://www.arXiv.org/abs/hep-ph/0201081}{{\tt hep-ph/0201081}}.

\bibitem{Baer:2001yy}
H.~Baer and J.~Ferrandis, ``{Supersymmetric SO(10) GUT models with Yukawa
  unification and a positive $\mu$ term},'' {\em Phys.Rev.Lett.} {\bf 87}
  (2001) 211803,
\href{http://www.arXiv.org/abs/hep-ph/0106352}{{\tt hep-ph/0106352}}.

\bibitem{Tobe:2003bc}
K.~Tobe and J.~D. Wells, ``{Revisiting top bottom tau Yukawa unification in
  supersymmetric grand unified theories},'' {\em Nucl.Phys.} {\bf B663} (2003)
  123--140,
\href{http://www.arXiv.org/abs/hep-ph/0301015}{{\tt hep-ph/0301015}}.

\bibitem{Baer:2009ie}
H.~Baer, S.~Kraml, and S.~Sekmen, ``{Is 'just-so' Higgs splitting needed for t
  - b - tau Yukawa unified SUSY GUTs?},'' {\em JHEP} {\bf 0909} (2009) 005,
\href{http://www.arXiv.org/abs/0908.0134}{{\tt 0908.0134}}.

\bibitem{Badziak:2011wm}
M.~Badziak, M.~Olechowski, and S.~Pokorski, ``{Yukawa unification in SO(10)
  with light sparticle spectrum},'' {\em JHEP} {\bf 1108} (2011) 147,
\href{http://www.arXiv.org/abs/1107.2764}{{\tt 1107.2764}}.

\bibitem{Anandakrishnan:2012tj}
A.~Anandakrishnan, S.~Raby, and A.~Wingerter, ``{Yukawa Unification Predictions
  for the LHC},'' {\em Phys.Rev.} {\bf D87} (2013) 055005,
\href{http://www.arXiv.org/abs/1212.0542}{{\tt 1212.0542}}.

\bibitem{Anandakrishnan:2013cwa}
A.~Anandakrishnan and S.~Raby, ``{Yukawa Unification Predictions with effective
  "Mirage" Mediation},'' {\em Phys.Rev.Lett.} {\bf 111} (2013) 211801,
\href{http://www.arXiv.org/abs/1303.5125}{{\tt 1303.5125}}.

\bibitem{Pierce:1996zz}
D.~M. Pierce, J.~A. Bagger, K.~T. Matchev, and R.-j. Zhang, ``{Precision
  corrections in the minimal supersymmetric standard model},'' {\em Nucl.Phys.}
  {\bf B491} (1997) 3--67,
\href{http://www.arXiv.org/abs/hep-ph/9606211}{{\tt hep-ph/9606211}}.

\bibitem{ATLAS-CONF-2013-061}
``{Search for strong production of supersymmetric particles in final states
  with missing transverse momentum and at least three b-jets using 20.1 fb−1
  of pp collisions at sqrt(s) = 8 TeV with the ATLAS Detector.},'' Tech. Rep.
  ATLAS-CONF-2013-061, CERN, Geneva, Jun, 2013.

\bibitem{CMS-PAS-SUS-13-004}
{\bf CMS Collaboration} Collaboration, ``{Search for supersymmetry using razor
  variables in events with b-jets in pp collisions at 8 TeV},'' Tech. Rep.
  CMS-PAS-SUS-13-004, CERN, Geneva, 2013.

\bibitem{Anandakrishnan:2013nca}
A.~Anandakrishnan, B.~C. Bryant, S.~Raby, and A.~Wingerter, ``{LHC
  Phenomenology of SO(10) Models with Yukawa Unification},'' {\em Phys.Rev.}
  {\bf D88} (2013) 075002,
\href{http://www.arXiv.org/abs/1307.7723}{{\tt 1307.7723}}.

\bibitem{Chatrchyan:2013wxa}
{\bf CMS Collaboration} Collaboration, S.~Chatrchyan {\em et al.}, ``{Search
  for gluino mediated bottom- and top-squark production in multijet final
  states in pp collisions at 8 TeV},''
\href{http://www.arXiv.org/abs/1305.2390}{{\tt 1305.2390}}.

\bibitem{Chatrchyan:2012paa}
{\bf CMS Collaboration} Collaboration, S.~Chatrchyan {\em et al.}, ``{Search
  for new physics in events with same-sign dileptons and $b$ jets in $pp$
  collisions at $\sqrt{s}=8$ TeV},'' {\em JHEP} {\bf 1303} (2013) 037,
\href{http://www.arXiv.org/abs/1212.6194}{{\tt 1212.6194}}.

\bibitem{Chatrchyan:2013lya}
{\bf CMS Collaboration} Collaboration, S.~Chatrchyan {\em et al.}, ``{Search
  for supersymmetry in hadronic final states with missing transverse energy
  using the variables $\alpha_T$ and b-quark multiplicity in pp collisions at
  $\sqrt{s}$ = 8 TeV},''
\href{http://www.arXiv.org/abs/1303.2985}{{\tt 1303.2985}}.

\bibitem{Drees:2013wra}
M.~Drees, H.~Dreiner, D.~Schmeier, J.~Tattersall, and J.~S. Kim, ``{CheckMATE:
  Confronting your Favourite New Physics Model with LHC Data},''
\href{http://www.arXiv.org/abs/1312.2591}{{\tt 1312.2591}}.

\bibitem{Muhlleitner:2003vg}
M.~Muhlleitner, A.~Djouadi, and Y.~Mambrini, ``{SDECAY: A Fortran code for the
  decays of the supersymmetric particles in the MSSM},'' {\em
  Comput.Phys.Commun.} {\bf 168} (2005) 46--70,
\href{http://www.arXiv.org/abs/hep-ph/0311167}{{\tt hep-ph/0311167}}.

\bibitem{Choi:2005ge}
K.~Choi, A.~Falkowski, H.~P. Nilles, and M.~Olechowski, ``{Soft supersymmetry
  breaking in KKLT flux compactification},'' {\em Nucl.Phys.} {\bf B718} (2005)
  113--133,
\href{http://www.arXiv.org/abs/hep-th/0503216}{{\tt hep-th/0503216}}.

\bibitem{Choi:2005hd}
K.~Choi, K.~S. Jeong, T.~Kobayashi, and K.-i. Okumura, ``{Little SUSY hierarchy
  in mixed modulus-anomaly mediation},'' {\em Phys.Lett.} {\bf B633} (2006)
  355--361,
\href{http://www.arXiv.org/abs/hep-ph/0508029}{{\tt hep-ph/0508029}}.

\bibitem{Kitano:2005wc}
R.~Kitano and Y.~Nomura, ``{A Solution to the supersymmetric fine-tuning
  problem within the MSSM},'' {\em Phys.Lett.} {\bf B631} (2005) 58--67,
\href{http://www.arXiv.org/abs/hep-ph/0509039}{{\tt hep-ph/0509039}}.

\bibitem{Lowen:2008fm}
V.~Lowen and H.~P. Nilles, ``{Mirage Pattern from the Heterotic String},'' {\em
  Phys.Rev.} {\bf D77} (2008) 106007,
\href{http://www.arXiv.org/abs/0802.1137}{{\tt 0802.1137}}.

\bibitem{Anandakrishnan:2013tqa}
A.~Anandakrishnan and K.~Sinha, ``{On the Viability of Thermal Well-Tempered
  Dark Matter in SUSY GUTs},''
\href{http://www.arXiv.org/abs/1310.7579}{{\tt 1310.7579}}.

\bibitem{Choi:2005uz}
K.~Choi, K.~S. Jeong, and K.-i. Okumura, ``{Phenomenology of mixed
  modulus-anomaly mediation in fluxed string compactifications and brane
  models},'' {\em JHEP} {\bf 0509} (2005) 039,
\href{http://www.arXiv.org/abs/hep-ph/0504037}{{\tt hep-ph/0504037}}.

\bibitem{Krippendorf:2013dqa}
S.~Krippendorf, H.~P. Nilles, M.~Ratz, and M.~W. Winkler, ``{Hidden SUSY from
  precision gauge unification},'' {\em Phys.Rev.} {\bf D88} (2013) 035022,
\href{http://www.arXiv.org/abs/1306.0574}{{\tt 1306.0574}}.

\bibitem{Ade:2013ktc}
{\bf Planck Collaboration} Collaboration, P.~Ade {\em et al.}, ``{Planck 2013
  results. I. Overview of products and scientific results},''
\href{http://www.arXiv.org/abs/1303.5062}{{\tt 1303.5062}}.

\bibitem{Akerib:2013tjd}
{\bf LUX Collaboration} Collaboration, D.~Akerib {\em et al.}, ``{First results
  from the LUX dark matter experiment at the Sanford Underground Research
  Facility},''
\href{http://www.arXiv.org/abs/1310.8214}{{\tt 1310.8214}}.

\bibitem{Raby:2009sf}
S.~Raby, M.~Ratz, and K.~Schmidt-Hoberg, ``{Precision gauge unification in the
  MSSM},'' {\em Phys.Lett.} {\bf B687} (2010) 342--348,
\href{http://www.arXiv.org/abs/0911.4249}{{\tt 0911.4249}}.

\bibitem{Cheng:1998hc}
H.-C. Cheng, B.~A. Dobrescu, and K.~T. Matchev, ``{Generic and chiral
  extensions of the supersymmetric standard model},'' {\em Nucl.Phys.} {\bf
  B543} (1999) 47--72,
\href{http://www.arXiv.org/abs/hep-ph/9811316}{{\tt hep-ph/9811316}}.

\bibitem{Chen:1996ap}
C.~Chen, M.~Drees, and J.~Gunion, ``{A Nonstandard string / SUSY scenario and
  its phenomenological implications},'' {\em Phys.Rev.} {\bf D55} (1997)
  330--347,
\href{http://www.arXiv.org/abs/hep-ph/9607421}{{\tt hep-ph/9607421}}.

\bibitem{deFavereau:2013fsa}
{\bf DELPHES 3} Collaboration, J.~de~Favereau {\em et al.}, ``{DELPHES 3, A
  modular framework for fast simulation of a generic collider experiment},''
  {\em JHEP} {\bf 1402} (2014) 057,
\href{http://www.arXiv.org/abs/1307.6346}{{\tt 1307.6346}}.

\bibitem{Cacciari:2011ma}
M.~Cacciari, G.~P. Salam, and G.~Soyez, ``{FastJet User Manual},'' {\em
  Eur.Phys.J.} {\bf C72} (2012) 1896,
\href{http://www.arXiv.org/abs/1111.6097}{{\tt 1111.6097}}.

\bibitem{Cacciari:2005hq}
M.~Cacciari and G.~P. Salam, ``{Dispelling the $N^{3}$ myth for the $k_t$
  jet-finder},'' {\em Phys.Lett.} {\bf B641} (2006) 57--61,
\href{http://www.arXiv.org/abs/hep-ph/0512210}{{\tt hep-ph/0512210}}.

\bibitem{Cacciari:2008gp}
M.~Cacciari, G.~P. Salam, and G.~Soyez, ``{The Anti-k(t) jet clustering
  algorithm},'' {\em JHEP} {\bf 0804} (2008) 063,
\href{http://www.arXiv.org/abs/0802.1189}{{\tt 0802.1189}}.

\bibitem{Dobbs:2001ck}
M.~Dobbs and J.~B. Hansen, ``{The HepMC C++ Monte Carlo event record for High
  Energy Physics},'' {\em Comput.Phys.Commun.} {\bf 134} (2001)
41--46.

\bibitem{Sjostrand:2007gs}
T.~Sjostrand, S.~Mrenna, and P.~Z. Skands, ``{A Brief Introduction to PYTHIA
  8.1},'' {\em Comput.Phys.Commun.} {\bf 178} (2008) 852--867,
\href{http://www.arXiv.org/abs/0710.3820}{{\tt 0710.3820}}.

\bibitem{Kramer:2012bx}
M.~Kramer, A.~Kulesza, R.~van~der Leeuw, M.~Mangano, S.~Padhi, {\em et al.},
  ``{Supersymmetry production cross sections in $pp$ collisions at $\sqrt{s}=7$
  TeV},''
\href{http://www.arXiv.org/abs/1206.2892}{{\tt 1206.2892}}.

\bibitem{ATLAS-CONF-2013-047}
{\bf ATLAS} Collaboration, ``Search for squarks and gluinos with the atlas
  detector in final states with jets and missing transverse momentum and 20.3
  fb$^{-1}$ of $\sqrt{s}=8$ tev proton-proton collision data.''
  \url{http://cds.cern.ch/record/1547563/files/ATLAS-CONF-2013-047.pdf}, May,
  2013.
\newblock ATLAS-CONF-2013-047.

\bibitem{Anandakrishnan:2014pva}
A.~Anandakrishnan and C.~S. Hill, ``{Beyond Simplified Models: Constraining
  Supersymmetry on Triangles},''
\href{http://www.arXiv.org/abs/1403.4294}{{\tt 1403.4294}}.

\bibitem{Chatrchyan:2014lfa}
{\bf CMS Collaboration} Collaboration, S.~Chatrchyan {\em et al.}, ``{Search
  for new physics in the multijet and missing transverse momentum final state
  in proton-proton collisions at $\sqrt{s}$ = 8 TeV},''
\href{http://www.arXiv.org/abs/1402.4770}{{\tt 1402.4770}}.

\end{thebibliography}\endgroup

\bibliographystyle{utphys}

\end{document}